\documentclass{pi2005}

\usepackage{amsmath,amsthm,amsfonts,amssymb,cite}

\slacs{.4ex}

\begin{document}
\title{Casimir energy, the cosmological constant and massive gravitons}
\authori{Remo Garattini\,\footnote{E-mail:Remo.Garattini@unibg.it}}
\addressi{Universit\`{a} degli Studi di Bergamo, Facolt\`{a} di Ingegneria,\\
Viale Marconi 5, 24044 Dalmine (Bergamo) ITALY\\and\\INFN - sezione di Milano, Via Celoria 16, Milan, Italy}

\authorii{}
\addressii{}
\authoriii{}    \addressiii{}
\authoriv{}     \addressiv{}
\authorv{}      \addressv{}
\authorvi{}     \addressvi{}
\headauthor{Remo Garattini}
\headtitle{Casimir energy, the cosmological constant and massive
gravitons}
\lastevenhead{Remo Garattini: Casimir energy, the cosmological
constant and massive gravitons}
\pacs{04.60.-m, 04.62.+v, 11.10.Gh}
\keywords{Cosmological Constant, Renormalization, Quantum Gravity}

\maketitle

\begin{abstract}
The cosmological constant appearing in the Wheeler-De Witt
equation is considered as an eigenvalue of the associated
Sturm-Liouville problem. A variational approach with Gaussian
trial wave functionals is used as a method to study such a
problem. We approximate the equation to one loop in a
Schwarzschild background and a zeta function regularization is
involved to handle with divergences. The regularization is closely
related to the subtraction procedure appearing in the computation
of Casimir energy in a curved background. A renormalization
procedure is introduced to remove the infinities together with a
renormalization group equation. The case of massive gravitons is
discussed.
\end{abstract}

\section{Introduction}

One of the most fascinating and unsolved problems of the
theoretical physics of our century is the cosmological constant.
Einstein introduced his cosmological constant $\Lambda_c$ in an
attempt to generalize his original field equations. The modified
field equations are
\begin{equation}
R_{\mu\nu}-\sfrac{1}{2}\,g_{\mu\nu}R+\Lambda_{c}g_{\mu\nu}=8\pi
GT_{\mu\nu}, \label{i0}
\end{equation}
where $\Lambda_{c}$ is the cosmological constant, $G$ is the
gravitational constant and $T_{\mu\nu}$ is the energy--momentum
tensor. By redefining
$$
T_{\mu\nu}^{\mathrm{tot}}\equiv T_{\mu\nu}-\frac{\Lambda_{c}}{8\pi
G}\,g_{\mu\nu}\,,
$$
one can regain the original form of the field equations
\begin{equation}
R_{\mu\nu}-\sfrac{1}{2}\,g_{\mu\nu}R=8\pi GT_{\mu\nu}^{tot}=8\pi
G\left(T_{\mu\nu}+T_{\mu\nu}^{\Lambda}\right), \label{i2}
\end{equation}
at the prize of introducing a vacuum energy density and vacuum
stress--energy tensor
$$
\rho_{\Lambda}=\frac{\Lambda_{c}}{8\pi G}\,,\quad
T_{\mu\nu}^{\Lambda}=-\rho_{\Lambda}g_{\mu\nu}\,.
$$
Alternatively, Eq.~(\ref{i0}) can be cast into the form,
$$
R_{\mu\nu}-\sfrac{1}{2}\,g_{\mu\nu}R+\Lambda_{\mathrm{eff}}g_{\mu\nu}=0\,,
$$
where we have included the contribution of the vacuum energy
density in the form $T_{\mu\nu}=-\langle\rho\rangle g_{\mu\nu}$.
In this case $\Lambda_{c}$ can be considered as the bare
cosmological constant
$$
\Lambda_{\mathrm{eff}}=8\pi G\rho_{\mathrm{eff}}= \Lambda_{c}+8\pi
G\langle\rho\rangle\,.
$$
Experimentally, we know that the effective energy density of the
universe $\rho_{\mathrm{eff}}$ is of the order
$10^{-47}\,\mathrm{GeV}^4$. A crude estimate of the Zero Point
Energy (ZPE) of some field of mass $m$ with a cutoff at the Planck
scale gives
\begin{equation}
E_{\mathrm{ZPE}}=\frac{1}{2}\int_0^{\Lambda_p}\frac{\D^3k}{\left(2\pi\right)^3}
\sqrt{k^2+m^2}\simeq\frac{\Lambda_p^4}{16\pi^2}\approx
10^{71}\,\mathrm{GeV}^4\,.
\label{zpe}
\end{equation}
This gives a difference of about 118 orders \cite{Lambda}. The
approach to quantization of general relativity based on the
following set of equations
\begin{equation}
\left[2\kappa G_{ijkl}\pi^{ij}\pi^{kl}-\frac{\sqrt{g}}{2\kappa}
\left(R-2\Lambda_c\right)\right]\Psi\left[g_{ij}\right]=0
\label{WDW}
\end{equation}
and
\begin{equation}
-2\nabla_i\pi^{ij}\Psi\left[ g_{ij}\right]=0\,,
\label{diff}
\end{equation}
where $R$ is the three-scalar curvature, $\Lambda_c$ is the bare
cosmological constant and $\kappa=8\pi G$, is known as Wheeler--De
Witt equation (WDW) \cite{DeWitt}. Eqs.~(\ref{WDW}) and
(\ref{diff}) describe the \textit{wave function of the universe}.
The WDW equation represents invariance under \textit{time}
reparametrization in an operatorial form, while Eq.~(\ref{diff})
represents invariance under diffeomorphism. $G_{ijkl}$ is the
\textit{supermetric} defined as
$$
G_{ijkl}=\frac{1}{2\sqrt{g}}\,(g_{ik}g_{jl}+g_{il}g_{jk}-g_{ij}g_{kl})\,.
$$
Note that the WDW equation can be cast into the form
$$
\left[2\kappa
G_{ijkl}\pi^{ij}\pi^{kl}-\frac{\sqrt{g}}{2\kappa}\,R\right]
\Psi\left[g_{ij}\right]=-\frac{\sqrt{g}}{\kappa}\,\Lambda_{c}\Psi\left[g_{ij}\right],
$$
which formally looks like an eigenvalue equation. In this paper,
we would like to use the Wheeler--De Witt (WDW) equation to
estimate $\langle\rho\rangle$. In particular, we will compute the
ZPE due to massive and massless gravitons propagating on the
Schwarzschild background. This choice is dictated by considering
that the Schwarzschild solution represents the only non-trivial
static spherical symmetric solution of the Vacuum Einstein
equations. Therefore, in this context the ZPE can be attributed
only to quantum fluctuations. The used method will be a
variational approach applied on gaussian wave functional. The rest
of the paper is structured as follows, in section \ref{p1}, we
show how to apply the variational approach to the Wheeler--De Witt
equation and we give some of the basic rules to solve such an
equation approximated to second order in metric perturbation, in
section \ref{p2}, we analyze the spin-2 operator or the operator
acting on transverse traceless tensors, in section \ref{p3} we use
the zeta function to regularize the divergences coming from the
evaluation of the ZPE for TT tensors and we write the
renormalization group equation, in section \ref{p4} we use the
same procedure of section \ref{p3}, but for massive gravitons. We
summarize and conclude in section \ref{p5}.

\section{The Wheeler\bmth{-}De Witt equation and the cosmological constant}
\label{p1}

The WDW equation (\ref{WDW}), written as an
eigenvalue equation, can be cast into the form%
\begin{equation}
\hat{\Lambda}_{\Sigma}\Psi\left[g_{ij}\right]=
-\Lambda^{\prime}(x)\Psi\left[g_{ij}\right],
\label{WDW1}
\end{equation}
where
$$ \left\{\begin{array}{l} \hat{\Lambda}_{\Sigma}=2\kappa
G_{ijkl}\pi^{ij}\pi^{kl}-\dfrac{\sqrt{g}}{2\kappa}R\,,\\[9pt]
\Lambda^{\prime}=\dfrac{\Lambda}{\kappa}\sqrt{g}\,.
\end{array}\right.
$$
We, now multiply Eq.~(\ref{WDW1}) by
$\Psi^{\ast}\left[g_{ij}\right]$ and we functionally integrate
over the three spatial metric $g_{ij}$, then after an integration
over the hypersurface $\Sigma$, one can formally re-write the WDW
equation as
\begin{equation}
\frac{1}{V}\,\frac{\int\mathcal{D}\left[g_{ij}\right]
\Psi^{\ast}\left[g_{ij}\right]
\int_{\Sigma}\D^3x\hat{\Lambda}_{\Sigma}\Psi\left[g_{ij}\right]}
{\int\mathcal{D}\left[g_{ij}\right]\Psi^{\ast}\left[g_{ij}\right]
\Psi\left[g_{ij}\right]}=\frac{1}{V}\,\frac{\left\langle\Psi\left\vert
\int_{\Sigma}\D^3x\hat{\Lambda}_{\Sigma}\right\vert
\Psi\right\rangle}{\left\langle\Psi|\Psi\right\rangle}=\Lambda^{\prime}\,.
\label{WDW2}
\end{equation}
The formal eigenvalue equation is a simple manipulation of
Eq.~(\ref{WDW}). However, we gain more information if we consider
a separation of the spatial part of the metric into a background
term, $\bar{g}_{ij}$, and a perturbation, $h_{ij}$,
$$
g_{ij}=\bar{g}_{ij}+h_{ij}\,.
$$
Thus eq.~(\ref{WDW2}) becomes
\begin{equation}
\frac{\left\langle\Psi\left\vert\int_{\Sigma}\D^{3}x
\left[\hat{\Lambda}_{\Sigma}^{(0)}+\hat{\Lambda}_{\Sigma}^{(1)}+
\hat{\Lambda}_{\Sigma}^{(2)}+\ldots\right]
\right\vert\Psi\right\rangle}{\left\langle\Psi|\Psi\right\rangle}=
\Lambda^{\prime}\Psi\left[g_{ij}\right],
\label{WDW3}
\end{equation}
where $\hat{\Lambda}_{\Sigma}^{(i)}$ represents the
$i^{\mathrm{th}}$ order of perturbation in $h_{ij}$. By observing
that the kinetic part of $\hat{\Lambda}_{\Sigma}$ is quadratic in
the momenta, we only need to expand the three-scalar curvature
$\int\D^{3}x\sqrt{g}R^{(3)}$ up to quadratic order and we get
\begin{equation}
\label{rexp}
\begin{array}{l}
\disty
\int_{\Sigma}\D^{3}x\sqrt{\bar{g}}\Bigl[-\sfrac{1}{4}\,h\triangle
h+\sfrac{1}{4}\,h^{li}\triangle
h_{li}-\sfrac{1}{2}\,h^{ij}\nabla_{l}\nabla_{i}h_{j}^{l}+\\[6pt]
\disty\qquad\qquad +\sfrac{1}{2}\,h\nabla_{l}\nabla_{i}h^{li}-
\sfrac{1}{2}\,h^{ij}R_{ia}h_{j}^{a}+\sfrac{1}{2}\,hR_{ij}h^{ij}+
\sfrac{1}{4}\,h\left(R^{(0)}\right)h\Bigr]\,,
\end{array}
\end{equation}
where $h$ is the trace of $h_{ij}$ and $R^{(0)}$ is the three
dimensional scalar curvature. To explicitly make calculations, we
need an orthogonal decomposition for both $\pi_{ij}$ and $h_{ij}$
to disentangle gauge modes from physical deformations. We define
the inner product
$$
\langle h,k\rangle:=\int_{\Sigma}\sqrt{g}G^{ijkl}h_{ij}(x)
k_{kl}(x)\,\D^{3}x\,,
$$
by means of the inverse WDW metric $G_{ijkl}$, to have a metric on
the space of deformations, i.e. a quadratic form on the tangent
space at $h_{ij}$, with
$$
G^{ijkl}=(g^{ik}g^{jl}+g^{il}g^{jk}-2g^{ij}g^{kl})\,.
$$
The inverse metric is defined on cotangent space and it assumes
the form
$$
\langle p,q\rangle:=\int_{\Sigma}\sqrt{g}G_{ijkl}p^{ij}(x)
q^{kl}(x)\,\D^{3}x\,,
$$
so that
$$
G^{ijnm}G_{nmkl}=\sfrac{1}{2}\left(\delta_k^i\delta_l^j+
\delta_l^i\delta_k^j\right).
$$
Note that in this scheme the ``inverse metric" is actually the WDW
metric defined on phase space. The desired decomposition on the
tangent space of 3-metric deformations
\cite{BergerEbin,York,MazurMottola,Vassilevich} is:
\begin{equation}
h_{ij}=\sfrac{1}{3}\,hg_{ij}+(L\xi)_{ij}+h_{ij}^{\bot}\,,
\label{p21a}
\end{equation}
where the operator $L$ maps $\xi_{i}$ into symmetric tracefree
tensors
\begin{equation}
(L\xi)_{ij}=\nabla_{i}\xi_{j}+\nabla_{j}\xi_{i}-\sfrac{2}{3}\,g_{ij}\left(\nabla\cdot\xi\right).
\end{equation}
Thus the inner product between three-geometries becomes
\begin{equation}
\langle
h,h\rangle:=\int_{\Sigma}\sqrt{g}G^{ijkl}h_{ij}(x)h_{kl}(x)\D^{3}x=
\int_{\Sigma}\sqrt{g}\left[-\sfrac{2}{3}\,h^{2}+(L\xi)^{ij}(L\xi)_{ij}+
h^{ij\bot}h_{ij}^{\bot}\right].
\label{p21b}
\end{equation}
With the orthogonal decomposition in hand we can define the trial
wave functional as
\begin{equation}
\Psi\left[h_{ij}\left(\overrightarrow{x}\right)\right]=
\mathcal{N}\Psi\left[h_{ij}^{\bot}\left(\overrightarrow{x}\right)\right]
\Psi\left[h_{ij}^{\Vert}\left(\overrightarrow{x}\right)\right]
\Psi\left[h_{ij}^{\mathrm{trace}}\left(\overrightarrow{x}\right)\right],
\label{twf}
\end{equation}
where
$$
\begin{array}{rcl}
\Psi\left[ h_{ij}^{\bot}\left(\overrightarrow{x}\right)\right]&=&
\exp\left\{-\frac{1}{4}\left\langle
hK^{-1}h\right\rangle_{x,y}^{\bot}\right\},\\[6pt]
\Psi\left[h_{ij}^{\Vert}\left(\overrightarrow{x}\right)\right]&=&
\exp\left\{-\frac{1}{4}\left\langle\left(L\xi\right)K^{-1}
\left(L\xi\right)\right\rangle_{x,y}^{\Vert}\right\},\\[6pt]
\Psi\left[h_{ij}^{\mathrm{trace}}\left(\overrightarrow{x}\right)\right]&=&
\exp\left\{-\frac{1}{4}\left\langle
hK^{-1}h\right\rangle_{x,y}^{\mathrm{trace}}\right\}.
\end{array}
$$
The symbol ``$\perp$" denotes the transverse-traceless tensor (TT)
(spin 2) of the perturbation, while the symbol ``$\Vert$" denotes
the longitudinal part (spin 1) of the perturbation. Finally, the
symbol ``trace" denotes the scalar part of the perturbation.
$\mathcal{N}$ is a normalization factor,
$\langle\cdot,\cdot\rangle_{x,y}$ denotes space integration and
$K^{-1}$ is the inverse ``\textit{propagator}". We will fix our
attention to the TT tensor sector of the perturbation representing
the graviton. Therefore, representation (\ref{twf}) reduces to
\begin{equation}
\Psi\left[h_{ij}\left(\overrightarrow{x}\right)\right]=
\mathcal{N}\exp\left\{-\frac{1}{4}\left\langle
hK^{-1}h\right\rangle_{x,y}^{\bot}\right\}.
\label{tt}
\end{equation}
Actually there is no reason to neglect longitudinal perturbations.
However, following the analysis of Mazur and Mottola of
Ref.~\cite{MazurMottola} on the perturbation decomposition, we can
discover that the relevant components can be restricted to the TT
modes and to the trace modes. Moreover, for certain backgrounds TT
tensors can be a source of instability as shown in
Refs.~\cite{Instability}. Even the trace part can be regarded as a
source of instability. Indeed this is usually termed
\textit{conformal} instability. The appearance of an instability
on the TT modes is known as non conformal instability. This means
that does not exist a gauge choice that can eliminate negative
modes. To proceed with Eq.~(\ref{WDW3}), we need to know the
action of some basic operators on $\Psi\left[h_{ij}\right]$. The
action of the operator $h_{ij}$ on
$|\Psi\rangle=\Psi\left[h_{ij}\right]$ is realized by
\cite{Variational}
$$
h_{ij}(x)|\Psi\rangle=h_{ij}\left(\overrightarrow{x}\right)
\Psi\left[h_{ij}\right].
$$
The action of the operator $\pi_{ij}$ on $|\Psi\rangle$, in
general, is
$$
\pi_{ij}(x)|\Psi\rangle=-\I\,\frac{\delta}{\delta h_{ij}
\left(\overrightarrow{x}\right)}\,\Psi\left[h_{ij}\right],
$$
while the inner product is defined by the functional integration:
$$
\left\langle\Psi_{1}\mid\Psi_{2}\right\rangle=
\int\left[\mathcal{D}h_{ij}\right]\Psi_{1}^{\ast}\left[h_{ij}\right]
\Psi_{2}\left[h_{kl}\right].
$$
We demand that
\begin{equation}
\frac{1}{V}\,\frac{\left\langle\Psi\left\vert\int_{\Sigma}\D^{3}x\hat{\Lambda}_{\Sigma}
\right\vert\Psi\right\rangle}{\left\langle\Psi|\Psi\right\rangle}=
\frac{1}{V}\,\frac{\int\mathcal{D}\left[g_{ij}\right]\Psi^{\ast}\left[h_{ij}\right]
\int_{\Sigma}\D^{3}x\hat{\Lambda}_{\Sigma}\Psi\left[h_{ij}\right]}
{\int\mathcal{D}\left[g_{ij}\right]\Psi^{\ast}\left[h_{ij}\right]\Psi\left[h_{ij}\right]}
\label{vareq}
\end{equation}
be stationary against arbitrary variations of
$\Psi\left[h_{ij}\right]$. Note that Eq.~(\ref{vareq}) can be
considered as the variational analog of a Sturm--Liouville problem
with the cosmological constant regarded as the associated
eigenvalue. Therefore the solution of Eq.~(\ref{WDW2}) corresponds
to the minimum of Eq.~(\ref{vareq}). The form of
$\left\langle\Psi\left\vert\hat{\Lambda}_{\Sigma}\right\vert\Psi\right\rangle$
can be computed with the help of the wave functional (\ref{tt})
and with the help of
$$
\left\{\begin{array}{l} \dfrac{\left\langle\Psi\left\vert
h_{ij}\left(\overrightarrow{x}\right)\right\vert\Psi\right\rangle}
{\left\langle\Psi|\Psi\right\rangle}=0\,,\\[9pt]
\dfrac{\left\langle\Psi\left\vert h_{ij}
\left(\overrightarrow{x}\right)h_{kl}
\left(\overrightarrow{y}\right)\right\vert\Psi\right\rangle}
{\left\langle\Psi|\Psi\right\rangle}=
K_{ijkl}\left(\overrightarrow{x},\overrightarrow{y}\right)\,.
\end{array}\right.
$$
Extracting the TT tensor contribution, we get
\begin{equation}
\hat{\Lambda}_{\Sigma}^{\bot}=\frac{1}{4V}\int_{\Sigma}\D^{3}x\sqrt{\bar{g}}
G^{ijkl}\left[2\kappa K^{-1\bot}(x,x)_{ijkl}+
\frac{1}{2\kappa}\left(\triangle_2\right)_{j}^{a}
K^{\bot}(x,x)_{iakl}\right].
\label{p22}
\end{equation}
The propagator $K^{\bot}(x,x)_{iakl}$ can be represented as
\begin{equation}
K^{\bot}\left(\overrightarrow{x},\overrightarrow{y}\right)_{iakl}:=
\sum_{\tau}\frac{h_{ia}^{(\tau)\bot}\left(\overrightarrow{x}\right)
h_{kl}^{(\tau)\bot}\left(\overrightarrow{y}\right)}
{2\lambda(\tau)}\,,
\label{proptt}
\end{equation}
where $h_{ia}^{(\tau)\bot}\left(\overrightarrow{x}\right)$ are the
eigenfunctions of $\triangle_2$. $\tau$ denotes a complete set of
indices and $\lambda(\tau)$ are a set of variational parameters to
be determined by the minimization of Eq.~(\ref{p22}). The
expectation value of $\hat{\Lambda}_{\Sigma}^{\bot}$ is easily
obtained by inserting the form of the propagator into
Eq.~(\ref{p22})
$$
\Lambda^{\prime}\left(\lambda_i\right)=\frac{1}{4}\sum_{\tau}
\sum_{i=1}^2\left[2\kappa\lambda_i(\tau)+\frac{\omega_i^2(\tau)}
{2\kappa\lambda_i(\tau)}\right].
$$
By minimizing with respect to the variational function
$\lambda_i(\tau)$, we obtain the total one loop energy density for
TT tensors
\begin{equation}
\Lambda\left(\lambda_i\right)=-\frac{\kappa}{4}\sum_{\tau}
\left[\sqrt{\omega_1^2(\tau)}+\sqrt{\omega_2^2(\tau)}\right].
\label{lambda1loop}
\end{equation}
The above expression makes sense only for $\omega_i^2(\tau)>0$.

\section{The transverse traceless (TT) spin 2 operator for the Schwarzschild
metric and the W.K.B. approximation}
\label{p2}

The spin-two operator for the Schwarzschild metric is defined by
\begin{equation}
\left(\triangle_2h^{\mathrm{TT}}\right)_i^j:=
-\left(\triangle_{T}h^{\mathrm{TT}}\right)_i^j+2\left(Rh^{\mathrm{TT}}\right)_i^j\,,
\label{spin2}
\end{equation}
where the transverse-traceless (TT) tensor for the quantum
fluctuation is obtained by the following decomposition
$$
h_i^j=h_i^j-\sfrac13\,\delta_i^jh+\sfrac13\,\delta_i^jh=
\left(h^{T}\right)_i^j+\sfrac13\,\delta_i^jh\,.
$$
This implies that $\left(h^{T}\right)_i^j\delta_j^i=0$. The
transversality condition is applied on $\left(h^T\right)_i^j$ and
becomes $\nabla_j\left(h^T\right)_i^j=0$. Thus
\begin{equation}
-\left(\triangle_{T}h^{\mathrm{TT}}\right)_i^j=
-\triangle_S\left(h^{\mathrm{TT}}\right)_i^j+
\frac{6}{r^2}\left(1-\frac{2MG}r\right),
\label{tlap}
\end{equation}
where $\triangle_S$ is the scalar curved Laplacian, whose form is
\begin{equation}
\triangle_S=\left(1-\frac{2MG}{r}\right) \frac{\D^2}{\D r^2}+
\left(\frac{2r-3MG}{r^2}\right)\frac{\D}{\D r}- \frac{L^2}{r^2}
\label{slap}
\end{equation}
and $R_j^a$ is the mixed Ricci tensor whose components are:
$$
R_i^a=\left\{-\frac{2MG}{r^3},\frac{MG}{r^3},\frac{MG}{r^3}\right\},
$$
This implies that the scalar curvature is traceless. We are
therefore led to study the following eigenvalue equation
\begin{equation}
\left(\triangle_2h^{\mathrm{TT}}\right)_i^j=\omega^2h_j^i\,,
\label{p31}
\end{equation}
where $\omega^2$ is the eigenvalue of the corresponding equation.
In doing so, we follow Regge and Wheeler in analyzing the equation
as modes of definite frequency, angular momentum and parity
\cite{ReggeWheeler}. In particular, our choice for the
three-dimensional gravitational perturbation is represented by
its even-parity form%
\begin{equation}
\left(h^{\mathrm{even}}\right)_j^i(r,\vartheta,\phi)=
\mathop{\mathrm{diag}}\left[H(r),K(r),L(r)\right]Y_{lm}(\vartheta,\phi)\,,
\label{pert}
\end{equation}
with
$$
\left\{\begin{array}{l}
H(r)=h_1^1(r)-\frac13\,h(r)\,,\\[4pt]
K(r)=h_2^2(r)-\frac13\,h(r)\,,\\[4pt]
L(r)=h_3^3(r)-\frac13\,h(r)\,.
\end{array}\right.
$$
From the transversality condition we obtain $h_2^2(r)=h_3^3(r)$.
Then $K(r)=L(r)$. For a generic value of the angular momentum $L$,
representation (\ref{pert}) joined to Eq.~(\ref{tlap}) lead to the
following system of PDE's
\begin{equation}
\left\{\begin{array}{l}
\left(-\triangle_S+\dfrac{6}{r^{2}}\left(1-\dfrac{2MG}{r}\right)-
\dfrac{4MG}{r^{3}}\right)H(r)=\omega_{1,l}^2H(r)\,\\[9pt]
\left(-\triangle_S+\dfrac{6}{r^{2}}\left(1-\dfrac{2MG}{r}\right)+
\dfrac{2MG}{r^{3}}\right)K(r)=\omega_{2,l}^2K(r)\,.
\end{array}\right.
\label{p33}
\end{equation}
Defining reduced fields
$$
H(r)=\frac{f_1(r)}{r}\,\quad K(r)=\frac{f_2(r)}r
$$
and passing to the proper geodesic distance from the
\textit{throat} of the bridge
\begin{equation}
\D x=\pm\frac{\D r}{\sqrt{1-\frac{2MG}r}}\,,
\label{throat}
\end{equation}
the system (\ref{p33}) becomes
\begin{equation}
\left\{\begin{array}{l}
\left[-\dfrac{\D^2}{\D x^2}+V_1(r)\right]f_1(x)=\omega_{1,l}^2f_1(x)\,,\\[9pt]
\left[-\dfrac{\D^2}{\D x^2}+V_2(r)\right]f_2(x)=
\omega_{2,l}^2f_2(x)
\end{array}\right.
\label{p34}
\end{equation}
with
$$
\left\{\begin{array}{l}
V_1(r)=\dfrac{l(l+1)}{r^2}+U_1(r)\,,\\[9pt]
V_2(r)=\dfrac{l(l+1)}{r^2}+U_2(r)\,,
\end{array}\right.
$$
where we have defined $r\equiv r(x)$ and
$$
\left\{\begin{array}{l}
U_1(r)=\left[\dfrac{6}{r^2}\left(1-\dfrac{2MG}{r}\right)-\dfrac{3MG}{r^3}\right],\\[9pt]
U_2(r)=\left[\dfrac{6}{r^2}\left(1-\dfrac{2MG}{r}\right)+\dfrac{3MG}{r^3}\right].
\end{array}\right.
$$
Note that
\begin{equation}
\left\{\begin{array}{lll}
U_1(r)\geq0\,,\quad&\mathrm{when}\quad& r\geq\frac52\,MG\,,\\[4pt]
U_1(r)<0\quad&\mathrm{when}\quad&2MG\leq r<\frac52\,MG\,,\\[4pt]
U_2(r)>0\quad&&\forall\,r\in\left[2MG,+\infty\right)\,.
\end{array}\right.
\label{negU}
\end{equation}
The functions $U_1(r)$ and $U_2(r)$ play the r\^{o}le of two
$r$-dependent effective masses $m_1^2(r)$ and $m_2^2(r)$,
respectively. In order to use the WKB approximation, we define two
$r$-dependent radial wave numbers
$k_1\left(r,l,\omega_{1,nl}\right)$ and
$k_2\left(r,l,\omega_{2,nl}\right)$
\begin{equation}
\left\{\begin{array}{l}
k_1^2\left(r,l,\omega_{1,nl}\right)=\omega_{1,nl}^2-\dfrac{l(l+1)}{r^2}-
m_1^2(r)\,,\\[9pt]
k_2^2\left(r,l,\omega_{2,nl}\right)=\omega_{2,nl}^2-\dfrac{l(l+1)}{r^2}-
m_2^2(r)
\end{array}\right.
\label{rwn}
\end{equation}
for $r\geq\frac52\,MG$. When $2MG\leq r<\frac52\,MG$,
$k_1^2\left(r,l,\omega_{1,nl}\right)$ becomes
\begin{equation}
k_1^2\left(r,l,\omega_{1,nl}\right)=\omega_{1,nl}^2-\frac{l(l+1)}{r^2}+
m_1^2(r)\,.
\label{rwn1}
\end{equation}

\section{One loop energy regularization and renormalization}
\label{p3}

In this section, we proceed to evaluate Eq.~(\ref{lambda1loop}).
The method is equivalent to the scattering phase shift method and
to the same method used to compute the entropy in the brick wall
model. We begin by counting the number of modes with frequency
less than $\omega_i$, $i=1,2$. This is given approximately by
\begin{equation}
\tilde{g}\left(\omega_i\right)=\sum_l\nu_i\left(l,\omega_i\right)
(2l+1)\,,
\label{p41}
\end{equation}
where $\nu_i\left(l,\omega_i\right)$, $i=1,2$ is the number of
nodes in the mode with $\left(l,\omega_i\right)$, such that
$\bigl(r\equiv r(x)\bigr)$
\begin{equation}
\nu_i\left(l,\omega_i\right)=\frac{1}{2\pi}\int_{-\infty}^{+\infty}
\D x\sqrt{k_i^2\left(r,l,\omega_i\right)}\,.
\label{p42}
\end{equation}
Here it is understood that the integration with respect to $x$ and
$l$ is taken over those values which satisfy
$k_i^2\left(r,l,\omega_i\right)\geq0$, $i=1,2$. With the help of
Eqs.~(\ref{p41}, \ref{p42}), we obtain the one loop total energy
for TT tensors
$$
\frac{1}{8\pi}\sum_{i=1}^2\int_{-\infty}^{+\infty}\D x
\left[\int_0^{+\infty}\omega_i\,\frac{\D\tilde{g}(\omega_i)}
{\D\omega_i}\,\D\omega_i\right].
$$
By extracting the energy density contributing to the cosmological
constant, we
get%
\begin{equation}
\label{tote1loop}
\begin{array}{l}
\Lambda=\Lambda_1+\Lambda_2=\rho_1+\rho_2=\\[6pt]
\disty\quad
=-\frac{\kappa}{16\pi^2}\int_0^{+\infty}\omega_1^2\sqrt{\omega_1^2-m_1^2(r)}\,\D\omega_1-
\frac{\kappa}{16\pi^2}\int_0^{+\infty}\omega_2^2\sqrt{\omega_2^2-m_2^2(r)}\,\D\omega_2\,,
\end{array}
\end{equation}
where we have included an additional $4\pi$ coming from the
angular integration. We use the zeta function regularization
method to compute the energy densities $\rho_1$ and $\rho_2$. Note
that this procedure is completely equivalent to the subtraction
procedure of the Casimir energy computation where the zero point
energy (ZPE) in different backgrounds with the same asymptotic
properties is involved. To this purpose, we introduce the
additional mass parameter $\mu$ in order to restore the correct
dimension for the regularized quantities. Such an arbitrary mass
scale emerges unavoidably in any regularization scheme. Then we
have
\begin{equation}
\rho_i(\varepsilon)=\frac{1}{16\pi^2}\,\mu^{2\varepsilon}
\int_0^{+\infty}\D\omega_i\frac{\omega_i^2}
{\left(\omega_i^2-m_i^2(r)\right)^{\varepsilon-1/2}}\,.
\label{zeta}
\end{equation}
The integration has to be meant in the range where
$\omega_i^2-m_i^2(r)\geq0$\footnote{Details of the calculation can
be found in the Appendix.}. One gets
\begin{equation}
\rho_i(\varepsilon)=\kappa\frac{m_i^2(r)}{256\pi^2}
\left[\frac{1}{\varepsilon}+\ln\left(\frac{\mu^2}{m_i^2(r)}\right)+
2\ln2-\frac12\right],
\label{zeta1}
\end{equation}
$i=1,2$. In order to renormalize the divergent ZPE, we write
$$
\Lambda=8\pi
G\bigl(\rho_1(\varepsilon)+\rho_2(\varepsilon)+\rho_1(\mu)+\rho_2(\mu)\bigr),
$$
where we have separated the divergent part from the finite part.
To handle with the divergent energy density we extract the
divergent part of $\Lambda$, in the limit
$\varepsilon\rightarrow0$ and we set
$$
\Lambda^{\mathrm{div}}=\frac{G}{32\pi\varepsilon}\left(m_1^4(r)+m_2^4(r)\right).
$$
Thus, the renormalization is performed via the absorption of the divergent
part into the re-definition of the bare classical constant $\Lambda$
$$
\Lambda\rightarrow\Lambda_0+\Lambda^{\mathrm{div}}\,.
$$

The remaining finite value for the cosmological constant reads
\begin{equation}
\begin{array}{l}
\dfrac{\Lambda_0}{8\pi G}=
\dfrac{1}{256\pi^2}\left\{m_1^4(r)\left[\ln\left(\dfrac{\mu^2}{\left\vert
m_1^2(r)\right\vert}\right)+2\ln2-\dfrac{1}{2}\right]\right.+\\[9pt]
\qquad\left.
+m_2^4(r)\left[\ln\left(\dfrac{\mu^2}{m_2^2(r)}\right)+
2\ln2-\dfrac{1}{2}\right]\right\}=\bigl(\rho_1(\mu)+\rho_2(\mu)\bigr)=
\rho_{\mathrm{eff}}^{\mathrm{TT}}(\mu,r)\,.
\end{array}
\label{lambda0}
\end{equation}

The quantity in Eq.~(\ref{lambda0}) depends on the arbitrary mass
scale $\mu$. It is appropriate to use the renormalization group
equation to eliminate such a dependence. To this aim, we impose
that \cite{RGeq}
\begin{equation}
\frac{1}{8\pi G}\mu\,
\frac{\partial\Lambda_0^{\mathrm{TT}}(\mu)}{\partial\mu}=
\mu\frac{\D}{\D\mu}\,\rho_{\mathrm{eff}}^{\mathrm{TT}}(\mu,r)\,.
\label{rg}
\end{equation}
Solving it we find that the renormalized constant $\Lambda_0$
should be treated as a running one in the sense that it varies
provided that the scale $\mu$ is changing
\begin{equation}
\Lambda_0(\mu,r) =\Lambda_0(\mu_0,r)+
\frac{G}{16\pi}\left(m_1^4(r)+m_2^4(r)\right)
\ln\frac{\mu}{\mu_0}\,.
\label{lambdamu}
\end{equation}
Substituting Eq.~(\ref{lambdamu}) into Eq.~(\ref{lambda0}) we find%
\begin{equation}
\begin{array}{rcl}
\dfrac{\Lambda_0(\mu_0,r)}{8\pi G}&=&-\dfrac{1}{256\pi^2}
\left\{m_1^4(r)\left[\ln\left(\dfrac{\left\vert
m_1^2(r)\right\vert}{\mu_0^2}\right)-2\ln2+\dfrac12\right]\right.+\\[12pt]
&&\left. +m_2^4(r)\left[\ln\left(\dfrac{m_2^2(r)}{\mu_0^2}\right)-
2\ln2+\dfrac{1}{2}\right]\right\}.
\end{array}
\label{lambdamu0}
\end{equation}
In order to fix the dependence of $\Lambda$ on $r$ and $M$, we
find the minimum of $\Lambda_0(\mu_0,r)$. To this aim, last
equation can be cast into the form\footnote{Recall
Eqs.~(\ref{rwn},\ref{rwn1}), showing a change of sign in
$m_1^2(r)$. Even if this is not the most appropriate notation to
indicate a change of sign in a quantity looking like a
``\textit{square effective mass}", this reveals useful in the zeta
function regularization and in the search for extrema.}
\begin{equation}
\frac{\Lambda_0(\mu_0,r)}{8\pi G}=-\frac{\mu_0^4}{256\pi^2}
\left\{x^2(r)\left[\ln\left(\frac{\vert x(r)\vert}{4}\right)+
\frac{1}{2}\right]+y^2(r)\left[\ln\left(\frac{y(r)}{4}\right)+
\frac{1}{2}\right]\right\},
\label{renrhoeff}
\end{equation}
where $x(r)=\pm m_1^2(r)/\mu_0^2$ and $y(r)=m_2^2(r)/\mu_0^2$. Now
we find the extrema of $\Lambda_0\left(\mu_0;x(r),y(r)\right)$ in
the range $\frac52\,MG\leq r$ and we get
\begin{equation}
\left\{\begin{array}{l}
x(r)=0\,,\\
y(r)=0\,,
\end{array}\right.
\label{rhomin1}
\end{equation}
which is never satisfied and
\begin{equation}
\left\{\begin{array}{l}
x(r)=4/\E\,,\\[4pt]
y(r)=4/\E\,,
\end{array}\right.\;\Longrightarrow\;
\left\{\begin{array}{l}
m_1^2(r)=4\mu_0^2/\E\,,\\[4pt]
m_2^2(r)=4\mu_0^2/\E\,,
\end{array}\right.
\label{rhomin2}
\end{equation}
which implies $M=0$ and $\bar{r}=\sqrt{3\E}/2\mu_0$. On the other
hand, in the range $2MG\leq r<\frac52\,MG$, we get again
$$
\left\{\begin{array}{l}
x(r)=0\,,\\
y(r)=0\,,
\end{array}\right.
$$
which has no solution and%
\begin{equation}
\left\{\begin{array}{l}
-m_1^2(r)=4\mu_0^2/\E\,,\\[4pt]
m_2^2(r)=4\mu_0^2/\E\,,
\end{array}\right.
\label{rhomin3}
\end{equation}
which implies
\begin{equation}
\left\{\begin{array}{l}
\bar{M}=4\mu_0^2\bar{r}^3/3\E G,,\\[4pt]
\bar{r}=\sqrt{6\E}/4\mu_0\,.
\end{array}\right.
\label{minE}
\end{equation}

Eq.~(\ref{renrhoeff}) evaluated on the minimum, now becomes
\begin{equation}
\Lambda_0\left(\bar{M},\bar{r}\right)=\frac{\mu_0^4G}{2\E^2\pi}\,,.
\label{lambdamin}
\end{equation}
It is interesting to note that thanks to the renormalization group
equation (\ref{rg}), we can directly compute $\Lambda_0$ at the
scale $\mu_0$ and only with the help of Eq.~(\ref{lambdamu}), we
have access at the scale $\mu$.

\section{One loop energy regularization and renormalization for massive gravitons}
\label{p4}

The question of massive gravitons is quite delicate. A tentative
to introduce a mass in the general framework has been done by
Boulware and Deser \cite{BoulwareDeser}, with the conclusion that
the theory is unstable and produces ghosts. However, at the
linearized level the Pauli--Fierz term \cite{PauliFierz}
\begin{equation}
S_{\mathrm{P.F.}}=\frac{m_g^2}{8\kappa}\int\D^4x\sqrt{-g}
\left[h^{\mu\nu}h_{\mu\nu}-h^2\right],
\label{mass}
\end{equation}
does not introduce ghosts. $m_g$ is the graviton mass. Following
Rubakov \cite{Rubakov}, the Pauli--Fierz term can be rewritten in
such a way to explicitly violate Lorentz symmetry, but to preserve
the three-dimensional Euclidean symmetry. In Minkowski space it
takes the form
\begin{equation}
\begin{array}{l}
\disty S_m=-\frac{1}{8\kappa}\int_{\mathcal{M}}\D^4x\sqrt{-g}
\bigl[m_0^2h^{00}h_{00}+2m_1^2h^{0i}h_{0i}-
m_2^2h^{ij}h_{ij}+\\[9pt]
\disty\hskip35mm + m_3^2h^{ii}h_{jj}-2m_4^2h^{00}h_{ii}\bigr]\,.
\end{array}
\label{SRubakov}
\end{equation}
A comparison between the massive action (\ref{SRubakov}) and the
Pauli--Fierz term shows that they can be set equal if we make the
following choice\footnote{See also Dubovski \cite{Dubovsky} for a
detailed discussion about the different choices of $m_1$, $m_2$,
$m_3$ and $m_4$}
$$
m_0^2=0\,,\quad m_1^2=m_2^2=m_3^2=m_4^2=m^2>0\,.
$$
If we fix the attention on the case%
\begin{equation}
m_0^2=m_1^2=m_3^2=m_4^2=0\,,\quad m_2^2=m^2>0\,,
\label{mg}
\end{equation}
we can see that the trace part disappears and we get
$$
S_m=\frac{m_g^2}{8\kappa}\int\D^4x\sqrt{-\hat{g}}\left[h^{ij}h_{ij}\right].
$$
The corresponding term in the linearized hamiltonian will be simply
$$
\mathcal{H}_m=-\frac{m_g^2}{8\kappa}\int\D^3xN\sqrt{\hat{g}}\left[h^{ij}h_{ij}\right].
$$
This means that Eq.~(\ref{spin2}), will be modified into
\begin{equation}
\left(\triangle_2h^{\mathrm{TT}}\right)_i^j:=-\left(\triangle_{T}h^{\mathrm{TT}}\right)_i^j+
2\left(Rh^{\mathrm{TT}}\right)_i^j+\left(m_g^2h^{\mathrm{TT}}\right)_i^j\,,
\label{spin3}
\end{equation}
Therefore, the square effective mass will be modified by adding
the term $m_g^2$. Note that, while $m_2^2(r)$ is constant in sign,
$m_1^2(r)$ is not. Indeed, for the critical value $\bar{r}=5MG/2$,
$m_1^2(\bar{r})=m_g^2$ and in the range $(2MG,5MG/2)$ for some
values of $m_g^2$, $m_1^2(\bar{r})$ can be negative. It is
interesting therefore concentrate in this range. To further
proceed, we observe that $m_1^2(r)$ and $m_2^2(r)$ can be recast
into a more suggestive and useful form, namely
$$
\left\{\begin{array}{l}
m_1^2(r)=m_g^2+U_1(r)=m_g^2+m_1^2(r,M)-m_2^2(r,M)\,,\\[4pt]
m_2^2(r)=m_g^2+U_2(r)=m_g^2+m_1^2(r,M)+m_2^2(r,M)\,,
\end{array}\right.
$$
where $m_1^2(r,M)\rightarrow0$ when $r\rightarrow\infty$ or
$r\rightarrow2MG$ and $m_2^2(r,M)=3MG/r^3$. Nevertheless, in the
above mentioned range $m_1^2(r,M)$ is negligible when compared
with $m_2^2(r,M)$. So, in a first approximation we can write%
$$
\left\{\begin{array}{l} m_1^2(r)\simeq m_g^2-m_2^2(r_0,M)
=m_g^2-m_0^2(M)\,,\\[4pt]
m_2^2(r)\simeq m_g^2+m_2^2(r_0,M)=m_g^2+m_0^2(M)\,,
\end{array}\right.
$$
where we have defined a parameter $r_0>2MG$ and
$m_0^2(M)=3MG/r_0^3$. The main reason for introducing a new
parameter resides in the fluctuation of the horizon that forbids
any kind of approach. Of course the quantum fluctuation must obey
the uncertainty relations. Thus, the
analogue of Eq.~(\ref{lambdamu0}) for massive gravitons becomes%
\begin{eqnarray}
\frac{\Lambda_0(\mu_0,r)}{8\pi G}&=&-\frac{1}{256\pi^2}
\left\{\left(m_g^2-m_0^2(M)\right)^2
\left[\ln\left(\frac{\left\vert m_g^2-m_0^2(M)\right\vert}
{\mu_0^2}\right)-2\ln2+\frac12\right]\right.+\nonumber\\
&&\left.
+\left(m_g^2+m_0^2(M)\right)^2\left[\ln\left(\frac{m_g^2+m_0^2(M)}
{\mu_0^2}\right)-2\ln2+\frac12\right]\right\}.
\label{lambdamu0a}
\end{eqnarray}
We can now discuss three cases:

\begin{enumerate}
\item $m_g^2\gg m_0^2(M)$,
\item $m_g^2=m_0^2(M)$,
\item $m_g^2\ll m_0^2(M)$.
\end{enumerate}

In case 1), we can rearrange Eq.~(\ref{lambdamu0a}) to obtain%
\begin{eqnarray*}
\frac{\Lambda_0(\mu_0,r)}{8\pi G}&=&-\frac{1}{256\pi^2}
\left\{2\left(m_g^4+m_0^4(M)\right)
\left[\ln\left(\frac{m_g^2}{4\mu_0^2}\right)+\frac12\right]\right.+\\
&&++\left(m_g^4+m_0^4(M)\right)
\left[\ln\left(1-\frac{m_0^2(M)}{m_g^2}\right)+
\ln\left(1+\frac{m_0^2(M}{m_g^2}\right)\right]+\\
&&\left.
+2m_g^2m_0^2(M)\left[\ln\left(1+\frac{m_0^2(M)}{m_g^2}\right)-
\ln\left(1-\frac{m_0^2(M)}{m_g^2}\right)\right]\right\}\simeq\\
&&\simeq-\frac{m_g^4}{256\pi^2}\left[2\ln\left(\frac{m_g^2}
{4\mu_0^2}\right)+1+3\left(\frac{m_0^2(M)}{m_g^2}\right)^2\right].
\end{eqnarray*}
The last term can be rearranged to give
$$
-\frac{m_g^4}{128\pi^2}\left[\ln\left(\frac{m_g^2}{4\mu_M^2}\right)
+\frac12\right],
$$
where we have introduced an intermediate scale defined by
\begin{equation}
\mu_M^2=\mu_0^2\exp\left(-\frac{3m_0^4(M)}{2m_g^4}\right).
\label{newscale}
\end{equation}
With the help of Eq.~(\ref{newscale}), the computation of the
minimum of $\Lambda_0^{\mathrm{TT}}$ is more simple. Indeed, if
$$
x=\frac{m_g^2}{4\mu_M^2},
$$
$\Lambda_0$ becomes
\begin{equation}
\Lambda_{0,M}(\mu_0,x)=-\frac{G\mu_M^4}{\pi}\,x^2 \left[\ln
x+\frac12\right].
\label{LambdansM}
\end{equation}
As a function of $x$, $\Lambda_{0,M}(\mu_0,x)$ vanishes for $x=0$
and $x=\E^{-1/2}$ and when
$x\in\left[0,\exp\left(-\frac12\right)\right] $,
$\Lambda_{0,M}^{\mathrm{TT}}(\mu_0,x)\geq0$. It has a maximum for
$$
\bar{x}=\frac{1}{\E}\;\Longleftrightarrow\;
m_g^2=\frac{4\mu_M^2}{\E}=\frac{4\mu_0^2}{\E}\,
\exp\left(-\frac{3m_0^4(M)}{2m_g^4}\right)
$$
and its value is
$$
\Lambda_{0,M}(\mu_0,\bar{x})
=\frac{G\mu_M^4}{2\pi\E^2}=\frac{G\mu_0^4}{2\pi\E^2}\,
\exp\left(-\frac{3m_0^4(M)}{m_g^4}\right)
$$
or
$$
\Lambda_{0,M}(\mu_0,\bar{x})=\frac{G}{32\pi}\,m_g^4
\exp\left(\frac{3m_0^4(M)}{m_g^4}\right).
$$

In case 2), Eq.~(\ref{lambdamu0a}) becomes
$$
\frac{\Lambda_0(\mu_0,r)}{8\pi G}\simeq
\frac{\Lambda_0(\mu_0)}{8\pi G}= -\frac{m_g^4}{128\pi^2}
\left[\ln\left(\frac{m_g^2}{4\mu_0^2}\right)+\frac12\right]
$$
or
$$
\frac{\Lambda_0(\mu_0)}{8\pi G}=
-\frac{m_0^4(M)}{128\pi^2}\left[\ln\left(\frac{m_0^2(M)}{4\mu_0^2}\right)
+\frac12\right].
$$
Again we define a dimensionless variable
$$
x=\frac{m_g^2}{4\mu_0^2}
$$
and we get%
\begin{equation}
\frac{\Lambda_{0,0}(\mu_0,x)}{8\pi G}=
-\frac{G\mu_0^4}{\pi}\,x^2\left[\ln x+\frac12\right].
\label{Lambdans0}
\end{equation}
The formal expression of Eq.~(\ref{Lambdans0}) is very close to
Eq.~(\ref{LambdansM}) and indeed the extrema are in the same
position of the scale variable $x$, even if the meaning of the
scale is here different. $\Lambda_{0,0}(\mu_0,x)$ vanishes for
$x=0$ and $x=4\E^{-1/2}$. In this range,
$\Lambda_{0,0}^{\mathrm{TT}}(\mu_0,x)\geq0$ and it has a minimum
located in
\begin{equation}
\bar{x}=\frac{1}{\E}\;\Longrightarrow\; m_g^2=\frac{4\mu_0^2}{\E}
\label{min}
\end{equation}
and
$$
\Lambda_{0,0}(\mu_0,\bar{x})=\frac{G\mu_0^4}{2\pi\E^2}
$$
or
$$
\Lambda_{0,0}(\mu_0,\bar{x})=
\frac{G}{32\pi}\,m_g^4=\frac{G}{32\pi}\,m_0^4(M)\,.
$$

Finally the case 3 ) leads to%
$$
\frac{\Lambda_0(\mu_0,r)}{8\pi G}\simeq
-\frac{m_0^4(M)}{256\pi^2}\left[2\ln\left(\frac{m_0^2(M)}{4\mu_0^2}\right)+
1+3\left(\frac{m_g^2}{m_0^2(M)}\right)^2\right].
$$
The last term can be rearranged to give
$$
-\frac{m_0^4(M)}{128\pi^2}\left[\ln\left(\frac{m_0^2(M)}{4\mu_m^2}\right)+
\frac12\right],
$$
where we have introduced another intermediate scale
$$
\mu_m^2=\mu_0^2\exp\left(-\frac{3m_g^4}{2m_0^4(M)}\right).
$$
By repeating the same procedure of previous cases, we define%
$$
x=\frac{m_0^2(M)}{4\mu_m^2}
$$
and we get
\begin{equation}
\Lambda_{0,m}(\mu_0,x)=-\frac{G\mu_m^4}{\pi}\,x^2 \left[\ln
x+\frac12\right].
\label{Lambdansm}
\end{equation}
Also this case has a maximum for
$$
\bar{x}=\frac{1}{\E}\;\Longrightarrow\;
m_0^2(M)=\frac{4\mu_m^2}{\E}=\frac{4\mu_0^2}{\E}
\exp\left(-\frac{3m_g^4}{2m_0^4(M)}\right).
$$
and
$$
\Lambda_{0,m}(\mu_0,\bar{x})=\frac{G\mu_m^4}{2\pi\E^2}=
\frac{G\mu_0^4}{2\pi\E^2}\exp\left(-\frac{3m_g^4}{m_0^4(M)}\right)
$$
or
$$
\Lambda_{0,M}(\mu_0,\bar{x})=\frac{G}{32\pi}\,m_0^4(M)
\exp\left(\frac{3m_g^4}{m_0^4(M)}\right).
$$

\noindent \textbf{Remark.} Note that in any case, the maximum of
$\Lambda$ corresponds to the minimum of the energy density.

\section{Summary and conclusions}
\label{p5}

In this paper, we have considered how to extract information on
the cosmological constant using the Wheeler--De Witt equation when
the graviton is massless and massive. In particular, by means of a
variational approach and a orthogonal decomposition of the modes,
we have studied the contribution of the transverse-traceless
tensors in a Schwarzschild background. The use of the zeta
function and a renormalization group equation have led to three
different cases:
$$
\left\{\begin{array}{l}
m_g^2\gg m_0^2(M)\\[4pt]
m_g^2=m_0^2(M)\\[4pt]
m_g^2\ll m_0^2(M)
\end{array}\right.\;\Longrightarrow\;
\left\{\begin{array}{l}
\Lambda_{0,M}(\mu_0,\bar{x})=G\mu_0^4/\left(2\pi\E^2\right)
\exp\left(-3m_0^4(M)/m_g^4\right)\\[6pt]
\Lambda_{0,0}(\mu_0,\bar{x})=G\mu_0^4/\left(2\pi\E^2\right)\\[6pt]
\Lambda_{0,m}(\mu_0,\bar{x})=G\mu_0^4/\left(2\pi\E^2\right)
\exp\left(-3m_g^4/m_0^4(M)\right)
\end{array}\right.
$$
As we can see, the case ``\textit{extreme}", where the graviton
mass is completely screened by the curvature ``\textit{mass}"
seems to have the biggest value. We recall that the highest is the
value of $\Lambda_0$ , the lowest is the value of the energy
density. However, the expression of the extreme case coincides
with the mass-less graviton discussed in section \ref{p3}. In that
paper, it is the curvature ``\textit{mass}" which plays the
r\^{o}le of the mass of the graviton and contributes to the
cosmological constant. So it appears that the gravitational field
in the background of the Schwarzschild metric generates a
``\textit{mass}" term, because of the curvature and this term
disappears when the Schwarzschild mass goes to zero. This leads to
the conclusion that fluctuations around Minkowski space do not
create a cosmological constant in absence of matter fields.
Nevertheless, this behavior works if we accept that near the
throat, vacuum fluctuations come into play forbidding to reach the
throat itself. If this is not the case and the throat can be
reached, then the curvature ``\textit{mass}" becomes completely
non-perturbative when the Schwarzschild mass $M\rightarrow0$. If
we choose to fix the renormalization point $\mu_0=m_p$, we obtain
approximately
$\Lambda_0^{\bot}(\bar{M},\bar{r})\simeq10^{37}\;\mathrm{GeV}^2$
which, in terms of energy density is in agreement with the
estimate of Eq.~(\ref{zpe}). Once fixed the scale $\mu_0$, we can
see what happens at the cosmological constant at the scale $\mu$,
by means of Eq.~(\ref{lambdamu}). What we see is that the
cosmological constant is vanishing at the sub-planckian scale
$\mu=m_p\E^{-1/4}$, but unfortunately is a scale which is very far
from the nowadays observations. Note that, because of the
condition (\ref{min}), the graviton mass becomes proportional to
the ``Planck mass", which is of the order $10^{16}$ GeV, while the
upper bound in eV is of the order $10^{-24}$\,--\,$10^{-29}$ eV
\cite{mg}. A quite curious thing comes on the estimate on the
``square graviton mass", which in this context is closely related
to the cosmological constant. Indeed, from Eq.~(\ref{min}) applied
on the square mass, we get
$$
m_g^2\propto\mu_0^2\simeq10^{32}\;\mathrm{GeV}^2=10^{50}\;\mathrm{eV}^2\,,
$$
while the experimental upper bound is of the order
$$
\left(
m_g^2\right)_{\mathrm{exp}}\propto10^{-48}-10^{-58}\;\mathrm{eV}^2\,,
$$
which gives a difference of about $10^{98}$\,--\,$10^{108}$
orders. This discrepancy strongly recall the difference of the
cosmological constant estimated at the Planck scale with that
measured in the space where we live. However, the analysis is not
complete. Indeed, we have studied the spectrum in a W.K.B.
approximation with the following condition
$k_i^2(r,l,\omega_i)\geq0$, $i=1,\,2$. Thus to complete the
analysis, we need to consider the possible existence of
nonconformal unstable modes, like the ones discovered in
Refs.~\cite{Instability}. If such an instability appears, this
does not mean that we have to reject the solution. In fact in
Ref.~\cite{Remo}, we have shown how to cure such a problem. In
that context, a\ model of ``space-time foam" has been introduced
in a large $N$ wormhole approach reproducing a correct decreasing
of the cosmological constant and simultaneously a stabilization of
the system under examination. Unfortunately in that approach a
renormalization scheme was missing and a W.K.B. approximation on
the wave function has been used to recover a Schr\"{o}dinger-like
equation. The possible next step is to repeat the scheme we have
adopted here in a large $N$ context, to recover the correct
vanishing behavior of the cosmological constant.

\vfill\eject

\appendix

\section{The zeta function regularization}
\label{app}

In this appendix, we report details on computation leading to
expression (\ref{zeta}). We begin with the following integral
\begin{equation}
\rho(\varepsilon)=\left\{\begin{array}{l} \disty
I_{+}=\mu^{2\varepsilon}\int_0^{+\infty}\D\omega\frac{\omega^2}
{\left(\omega^2+m^2(r)\right)^{\varepsilon-1/2}}\,,\\[9pt]
\disty
I_{-}=\mu^{2\varepsilon}\int_0^{+\infty}\D\omega\frac{\omega^2}
{\left(\omega^2-m^2(r)\right)}^{\varepsilon-1/2}\,,
\end{array}
\right.
\label{rho}
\end{equation}
with $m^2(r)>0$.

\subsection{\bmth{I_+} computation}
\label{app1}

If we define $t=\omega/\sqrt{m^2(r)}$, the integral $I_{+}$ in
Eq.~(\ref{rho}) becomes
$$
\begin{array}{l}
\disty \rho(\varepsilon)=\mu^{2\varepsilon}m^{4-2\varepsilon}(r)
\int_0^{+\infty}\D t
\frac{t^2}{\left(t^2+1\right)^{\varepsilon-1/2}}=
\frac12\,\mu^{2\varepsilon}m^{4-2\varepsilon}(r)
B\left(\frac32,\varepsilon-2\right)\,,\\[9pt]
\disty \frac12\,\mu^{2\varepsilon}m^{4-2\varepsilon}(r)
\frac{\Gamma\left(\frac32\right)\Gamma(\varepsilon-2)}
{\Gamma\left(\varepsilon-\frac12\right)}=
\frac{\sqrt{\pi}}4\,m^4(r)\left(\frac{\mu^2}{m^2(r)}\right)^{\varepsilon}
\frac{\Gamma(\varepsilon-2)}{\Gamma\left(\varepsilon-\frac{1}{2}\right)},
\end{array}
$$
where we have used the following identities involving the beta
function
$$
B(x,y)=2\int_0^{+\infty}\D t
\frac{t^{2x-1}}{\left(t^2+1\right)^{x+y}}\,,\quad
\mathop{\mathrm{Re}}x>0\,,\quad \mathop{\mathrm{Re}}y>0
$$
related to the gamma function by means of
$$
B(x,y) =\frac{\Gamma(x)\Gamma(y)}{\Gamma(x+y)}\,.
$$
Taking into account the following relations for the
$\Gamma$-function
\begin{equation}
\begin{array}{rcl}
\Gamma(\varepsilon-2)&=&\dfrac{\Gamma(1+\varepsilon)}{\varepsilon(\varepsilon-1)(\varepsilon-2)}\,,\\[9pt]
\Gamma(\varepsilon-\sfrac12)&=&\disty\frac{\Gamma(\varepsilon+\frac{1}{2})}
{\varepsilon-\frac{1}{2}}
\end{array}
\label{gamma}
\end{equation}
and the expansion for small $\varepsilon$
\begin{eqnarray*}
\Gamma(1+\varepsilon)&=&1-\gamma\varepsilon+O\left(\varepsilon^2\right),\\
\Gamma(\varepsilon+\sfrac12)&=&\Gamma(\sfrac12)-
\varepsilon\Gamma(\sfrac12)(\gamma+2\ln2)+O\left(\varepsilon^2\right),\\
x^{\varepsilon}&=&1+\varepsilon\ln x+O\left(\varepsilon^2\right),
\end{eqnarray*}
where $\gamma$ is the Euler's constant, we find
$$
\rho(\varepsilon)=-\frac{m^4(r)}{16}\left[\frac{1}{\varepsilon}+
\ln\left(\frac{\mu^2}{m^2(r)}\right)+2\ln2-\frac{1}{2}\right].
$$

\subsection{\bmth{I_{-}} computation}
\label{app2}

If we define $t=\omega/\sqrt{m^2(r)}$, the integral $I_{-}$ in
Eq.~(\ref{rho}) becomes
$$
\begin{array}{l}
\disty \rho(\varepsilon)=\mu^{2\varepsilon}m^{4-2\varepsilon}(r)
\int_0^{+\infty}\D
t\frac{t^2}{\left(t^2-1\right)^{\varepsilon-1/2}}=
\frac{1}{2}\,\mu^{2\varepsilon}m^{4-2\varepsilon}(r)
B\left(\varepsilon-2,\frac{3}{2}-\varepsilon\right),\\[12pt]
\disty \frac{1}{2}\,\mu^{2\varepsilon}m^{4-2\varepsilon}(r)
\frac{\Gamma(\frac{3}{2}-\varepsilon)\Gamma(\varepsilon-2)}
{\Gamma(-\frac12)}=
-\frac{1}{4\sqrt{\pi}}\,m^4(r)\left(\frac{\mu^2}{m^2(r)}\right)^{\varepsilon}
\Gamma\left(\frac{3}{2}-\varepsilon\right)\Gamma(\varepsilon-2),
\end{array}
$$
where we have used the following identity involving the beta
function
$$
\begin{array}{c}
\disty \frac{1}{p}\,B\left(1-\nu-\frac{\mu}{p},\nu\right)=
\int_1^{+\infty}\D t\,t^{\mu-1}\left(t^{p}-1\right)^{\nu-1}\\[9pt]
p>0\,,\quad \mathop{\mathrm{Re}}\nu>0\,,\quad
\mathop{\mathrm{Re}}\mu<p-p\mathop{\mathrm{Re}}\nu
\end{array}
$$
and the reflection formula
$$
\Gamma(z)\Gamma(1-z)=-z\Gamma(-z)\Gamma(z)\,.
$$
From the first of Eqs.~(\ref{gamma}) and from the expansion for
small $\varepsilon$
\begin{eqnarray*}
\Gamma(\sfrac{3}{2}-\varepsilon)&=&\Gamma(\sfrac{3}{2})
\bigl(1-\varepsilon(-\gamma-2\ln2+2)\bigr)+O\left(\varepsilon^2\right)\\
x^{\varepsilon}&=&1+\varepsilon\ln x+O\left(\varepsilon^2\right),
\end{eqnarray*}
we find
$$
\rho(\varepsilon)=-\frac{m^4(r)}{16}
\left[\frac{1}{\varepsilon}+\ln\left(\frac{\mu^2}{m^2(r)}\right)+
2\ln2-\frac12\right].
$$


\begin{thebibliography}{99}
\bibitem{Lambda}
For a pioneering review on this problem see S. Weinberg: Rev. Mod.
Phys. \textbf{61} (1989) 1. For more recent and detailed reviews
see V. Sahni and A. Starobinsky: Int. J. Mod. Phys. D \textbf{9}
(2000) 373, astro-ph/9904398; N. Straumann:
gr-qc/0208027; T.Padmanabhan: Phys.Rept. \textbf{380} (2003) 235;
hep-th/0212290.
\bibitem{DeWitt}
B.S. DeWitt: Phys. Rev. \textbf{160} (1967) 1113.
\bibitem{BergerEbin}
M. Berger and D. Ebin: J. Diff. Geom. \textbf{3} (1969) 379.
\bibitem{York}
J.W. York Jr.: J. Math. Phys. \textbf{14} (1973) 4; Ann. Inst.
Henri Poincar\'{e} A \textbf{21} (1974) 319.
\bibitem{MazurMottola}
P.O. Mazur and E. Mottola: Nucl. Phys. B \textbf{341} (1990) 187.
\bibitem{Vassilevich}
D.V. Vassilevich: Int. J. Mod. Phys. A \textbf{8} (1993) 1637;\\
D.V. Vassilevich: Phys. Rev. D \textbf{52} (1995) 999;
gr-qc/9411036.
\bibitem{Instability}
D.J. Gross, M.J. Perry and L.G. Yaffe: Phys. Rev. D \textbf{25}
(1982) 330;\\
B. Allen: Phys. Rev. D \textbf{30} (1984) 1153;\\
E. Witten: Nucl. Phys. B \textbf{195} (1982) 481;\\
P. Ginsparg and M.J. Perry: Nucl. Phys. B \textbf{222} (1983) 245;\\
R.E. Young: Phys. Rev. D \textbf{28} (1983) 2436;\\
R.E. Young: Phys. Rev. D \textbf{28} (1983) 2420;\\
S.W. Hawking and D.N. Page: Commun. Math. Phys. \textbf{87} (1983) 577;\\
R. Gregory and R. Laflamme: Phys. Rev. D \textbf{37} (1988) 305;\\
R. Garattini: Int. J. Mod. Phys. A \textbf{14} (1999) 2905;
gr-qc/9805096;\\
E Elizalde, S Nojiri and S.D. Odintsov: Phys. Rev. D \textbf{59}
(1999) 061501; hep-th 9901026\\
M.S. Volkov and A. Wipf: Nucl. Phys. B \textbf{582} (2000) 313;
hep-th/0003081;\\
R. Garattini: Class. Quant. Grav. \textbf{17} (2000) 3335;
gr-qc/0006076;\\
T. Prestidge: Phys. Rev. D \textbf{61} (2000) 084002;
hep-th/9907163;\\
S.S. Gubser and I. Mitra:
hep-th/0009126;\\
R. Garattini: Class. Quant. Grav. \textbf{18} (2001) 571;
gr-qc/0012078;\\
S.S. Gubser and I. Mitra: JHEP \textbf{8} (2001) 18;\\
J.P. Gregory and S.F. Ross: Phys. Rev. D \textbf{64} (2001)
124006; hep-th/0106220;\\
H.S. Reall: Phys. Rev. D \textbf{64} (2001) 044005;
hep-th/0104071;\\
G. Gibbons and S.A. Hartnoll: Phys. Rev. D \textbf{66} (2001)
064024; hep-th/0206202.
\bibitem{Variational}
A.K. Kerman and D. Vautherin: Ann. Phys. \textbf{192} (1989) 408;\\
J.M. Cornwall, R. Jackiw and E. Tomboulis: Phys. Rev. D \textbf{8}
(1974) 2428;\\
R. Jackiw: in: \textit{S\'{e}minaire de Math\'{e}matiques
Sup\'{e}rieures}, Montr\'{e}al, Qu\'{e}bec, Canada, June 1988,
Notes by P. de Sousa Gerbert;\\
M. Consoli and G. Preparata: Phys. Lett. B \textbf{154} (1985)
411.
\bibitem{ReggeWheeler}
T. Regge and J.A. Wheeler: Phys. Rev. \textbf{108} (1975) 1063.
\bibitem{RGeq}
J. Perez-Mercader and S.D. Odintsov: Int. J. Mod. Phys. D
\textbf{1} (1992) 401;\\
I.O. Cherednikov: Acta Physica Slovaca \textbf{52} (2002) 221;
I.O. Cherednikov: Acta Phys. Polon. B \textbf{35} (2004) 1607;\\
M. Bordag, U. Mohideen and V.M. Mostepanenko: Phys. Rep.
\textbf{353} (2001) 1. R. Garattini, \textsl{TSPU Vestnik} \textbf{44} \textbf{N7}, 72 (2004); gr-qc/0409016 .\\
Inclusion of non-perturbative effects, namely beyond one-loop, in
de Sitter Quantum Gravity have been discussed in S. Falkenberg and
S.D. Odintsov: Int. J. Mod. Phys. A \textbf{13} (1998) 607; hep-th
9612019.
\bibitem{BoulwareDeser}
D.G. Boulware and S. Deser: Phys. Rev. D \textbf{12} (1972) 3368.
\bibitem{PauliFierz}
M. Fierz and W. Pauli: Proc. Roy. Soc. Lond. A \textbf{173} (1939)
211.
\bibitem{Rubakov}
V.A. Rubakov:
hep-th/0407104.
\bibitem{Dubovsky}
S.L. Dubovsky:
hep-th/0409124.
\bibitem{mg}
A.S. Goldhaber and M.M. Nieto: Phys. Rev. D \textbf{9} (1974)
1119;\\
S.L. Larson and W.A. Hiscock: Phys. Rev. D \textbf{61} (2000)
104008; gr-qc/9912102.
\bibitem{Remo}
R. Garattini: Int. J. Mod. Phys. D \textbf{4} (2002) 635;
gr-qc/0003090.
\bibitem{GR}
I.S. Gradshteyn and I.M. Ryzhik: \textit{Table of Integrals,
Series, and Products}. (corrected and enlarged edition), edited by
A. Jeffrey, Academic Press, Inc.
\end{thebibliography}
\end{document}